\documentclass[twocolumn]{aastex701}
\usepackage{multirow}
\begin{document}

\title{RRAT~J2325--0530: A Rotating Radio Transient with an Atypical Waiting-time Distribution}

\correspondingauthor{Shi-Jie Gao, Xiang-Dong Li \& Zhen Yan}

\author[orcid=0000-0002-0822-0337,sname='Gao']{Shi-Jie Gao}
\affiliation{School of Astronomy and Space Science, Nanjing University, Nanjing, 210023, People's Republic of China}
\affiliation{Key Laboratory of Modern Astronomy and Astrophysics, Nanjing University, Ministry of Education, Nanjing, 210023, People's Republic of China}
\email[show]{gaosj@nju.edu.cn}

\author[orcid=0000-0002-0584-8145,sname='Li']{Xiang-Dong Li}
\affiliation{School of Astronomy and Space Science, Nanjing University, Nanjing, 210023, People's Republic of China}
\affiliation{Key Laboratory of Modern Astronomy and Astrophysics, Nanjing University, Ministry of Education, Nanjing, 210023, People's Republic of China}
\email[show]{lixd@nju.edu.cn}

\author[orcid=0000-0002-9322-9319,sname='Yan']{Zhen Yan}
\affiliation{Shanghai Astronomical Observatory, Chinese Academy of Sciences, Shanghai 200030, People's Republic of China}
\affiliation{School of Astronomy and Space Science, University of Chinese Academy of Sciences, Beijing 100049, People's Republic of China}
\email[show]{yanzhen@shao.ac.cn}

\author[orcid=0000-0001-5684-0103,sname='Shao']{Yi-Xuan Shao}
\affiliation{School of Astronomy and Space Science, Nanjing University, Nanjing, 210023, People's Republic of China}
\affiliation{Key Laboratory of Modern Astronomy and Astrophysics, Nanjing University, Ministry of Education, Nanjing, 210023, People's Republic of China}
\email[hide]{yixuan@smail.nju.edu.cn}

\author[orcid=0000-0002-5683-822X,sname='Zhou']{Ping Zhou}
\affiliation{School of Astronomy and Space Science, Nanjing University, Nanjing, 210023, People's Republic of China}
\affiliation{Key Laboratory of Modern Astronomy and Astrophysics, Nanjing University, Ministry of Education, Nanjing, 210023, People's Republic of China}
\email[hide]{pingzhou@nju.edu.cn}

\begin{abstract}

We present 1.25~GHz observations of the rotating radio transient (RRAT) J2325--0530, conducted with the Five-hundred-meter Aperture Spherical radio Telescope (FAST). Approximately 60\% of detected single pulses occur in clusters of 2 to 5 consecutive bursts. Consequently, the waiting-time distribution between successive single pulses exhibits a pronounced excess at one rotation period, deviating from the exponential distribution expected for a Poisson process.
After grouping consecutive bursts into single emission events, the recalculated waiting-time distribution is well described by a Weibull distribution with a shape parameter $k \gtrsim 1$. Monte Carlo simulations incorporating both intrinsic burst on-windows and rotational modulation successfully reproduce the observed one-rotation excess.
These results suggest that RRAT~J2325--0530 emits through a quasi-random process with on-windows slightly longer than its spin period, modulated by the the rotation of emission beam. Additionally, the polarization position angle shows complex behavior that cannot be fully described by the standard rotating vector model, and several pulses exhibit quasi-periodic micro-structures. Taken together, these features indicate complex magnetospheric dynamics underlying the sporadic emission behavior of RRAT~J2325--0530.
\end{abstract}


\section{Introduction} 
Rotating radio transients (RRATs) are a subclass of pulsar characterized by irregular and sporadic single pulses. They were first discovered through reanalyzing data from the Parkes Multibeam Pulsar Survey using single-pulse search techniques \citep{McLaughlin+2006}. To date, over 100 RRATs have been discovered\footnote{See the online RRATalog via \url{https://rratalog.github.io/rratalog/}.}.
The definition of RRATs remains somewhat unclear. Observationally, RRATs are typically defined as pulsars that are more easily detected through single-pulse searches than via standard periodicity-based methods \citep{Keane+2011,Cui+2017,Zhou+2023}.
A more physically motivated definition interprets RRATs as pulsars with extremely high nulling fractions, such that their emission ``on'' windows are shorter than a single rotation period \citep{Wang+2007,Burke-Spolaor+2010,Burke-Spolaor+2013}.
In this framework, RRAT are considered to represent an evolutionary phase of pulsars, in which increasing nulling behavior from the decay of magnetic field or the increase of spin period. Regular pulsars are thought to evolve through increasingly nulling states, ultimately reaching the extreme nulling regime characteristic of RRATs \citep{Burke-Spolaor+2010}.

RRAT~J2325--0530 was discovered in the 350~MHz drift-scan pulsar survey conducted with the Robert C. Byrd Green Bank Telescope \citep{Karako-Argaman+2015}. Follow-up observations across a broad band from 25 MHz to 1.4 GHz have revealed a period of $P=0.868735115026(9)~{\rm s}$, a period derivation of $\dot P\simeq 1.03\times10^{-15}~{\rm s~s^{-1}}$ and a dispersion measure of DM$=14.966\pm 0.007~{\rm pc~cm^{-3}}$ \citep{Taylor+2012,Karako-Argaman+2015,Meyers+2019,Kravtsov+2022}. The burst rates have been estimated as $43\pm 5~{\rm hr^{-1}}$ at 1.4~GHz with the Parkes radio telescope (Murriyang) and $73\pm7~{\rm hr^{-1}}$ at 154~MHz with the Murchison Widefield Array (MWA) \citep{Meyers+2019}.

Waiting time is the interval between two adjacent pulses, which is a key diagnostic for understanding emission processes in  both RRATs and fast radio bursts (FRBs). Analyzing waiting-time distributions provides crucial insights into burst clustering and underlying physical mechanisms, help distinguish between purely random processes and those governed by more complex dynamics. In a Poisson process, where pulses occur randomly and independently, the waiting-time distribution follows an exponential form. This behavior has been observed in some RRATs, including RRAT~J0139+3336 \citep{Xie+2022}, RRAT~J0628+0909 \citep{Hsu+2023}, RRAT~J1913+1330 \citep{Shapiro-Albert+2018,Zhong+2024} and RRAT~J1918--0449 \citep{Chen+2022}. In contrast, repeating FRBs exhibit non-Poissonian waiting-time distribution. Some sources are described by a Weibull distribution with shape parameters indicating burst clustering \citep{Oppermann+2018}, while others are better modeled by bimodal log-normal distributions with peaks at both millisecond and second timescales, which may reflect intrinsic source activity or propagation effects \citep{Li+2021,Xu+2022,Zhang+2023,Nimmo+2023,Xiao+2024,Zhou+2025}.

Previous observations of RRAT~J2325--0530 with Parkes and MWA suggested that its emission follows a Poisson-like process, with a waiting-time distribution described by an exponential model \citep{Meyers+2019}. However, our new observations with the Five-hundred-meter Aperture Spherical radio Telescope (FAST, \citealt{Nan+2006,Nan+2008}) reveal a significant excess at waiting times equal to one rotation period, inconsistent with the exponential form expected from a Poisson process. A similar excess appears to be present in the MWA data from \citet{Meyers+2019} (see their Fig.~8), but it was not emphasized likely due to the limited sample size and the use of coarse binning ($\sim 10$ rotations), which obscured the deviation from the exponential trend. In this work, we present our high-sensitivity FAST observations of RRAT~J2325--0530. In \autoref{sec:obs}, we describe the observing setup and data reduction. \autoref{sec:res} presents polarimetric analysis and the waiting-time statistics. In \autoref{sec:sum}, we discuss implications for the nature and evolution of RRATs and summarize our findings.

\begin{table*}[htb]
    \centering
    \caption{Observation log for RRAT~J2325--0530. Columns list the observation ID, UTC start time, MJD, effective integration length (excluding calibration intervals), observed RM (RM$_{\rm obs}$), ionospheric RM contribution (RM$_{\rm ion}$), interstellar RM (RM$_{\rm ISM}$), total number of detected single pulses ($N_{\rm SP}$), pulse rate, and multiplicity distribution of consecutive pulse groups (counts of groups with 1, 2, 3, 4 and 5 pulses).\label{tab:obs}}
    \begin{tabular}{cccccccccc}
    \hline
    Obs. & UTC Start Time& MJD & Length & RM$_{\rm obs}$&RM$_{\rm ion}$&RM$_{\rm ISM}$ & $N_{\rm SP}$ & Rate &Pulse Groups\\
          && (d) & (s)  & ($\rm rad\,m^{-2}$)& ($\rm rad\,m^{-2}$)& ($\rm rad\,m^{-2}$) &  & ($\rm hr^{-1}$) &(1, 2, 3, 4, 5)\\
    \hline
    1 & 2021-10-25T13:25:27&59512.56597& 3000  & $4.08\pm0.57$ &$1.01\pm0.10$&$3.07\pm0.58$&164&196.8&69, 29,  5, 3, 2\\
    2 & 2024-12-28T09:07:32&60672.38958& 3465  & $14.21\pm0.63$ &$10.92\pm0.10$&$3.29\pm0.64$&211&219.2&87, 26, 20, 3, 0\\
    3 & 2025-01-14T07:06:38&60689.30278& 3465  & $8.45\pm0.16$ &$8.11\pm0.10$&$0.34\pm0.19$&254&263.9&99, 27, 20, 9, 1\\
    4 & 2025-02-23T04:54:18&60729.21111& 3520  & $13.31\pm0.76$&$10.56\pm0.10$&$2.75\pm 0.77$&207&211.7&94, 22, 16, 4, 1\\
    \hline
    \end{tabular}
\end{table*}

\begin{figure*}
    \centering
    \includegraphics[width=\linewidth]{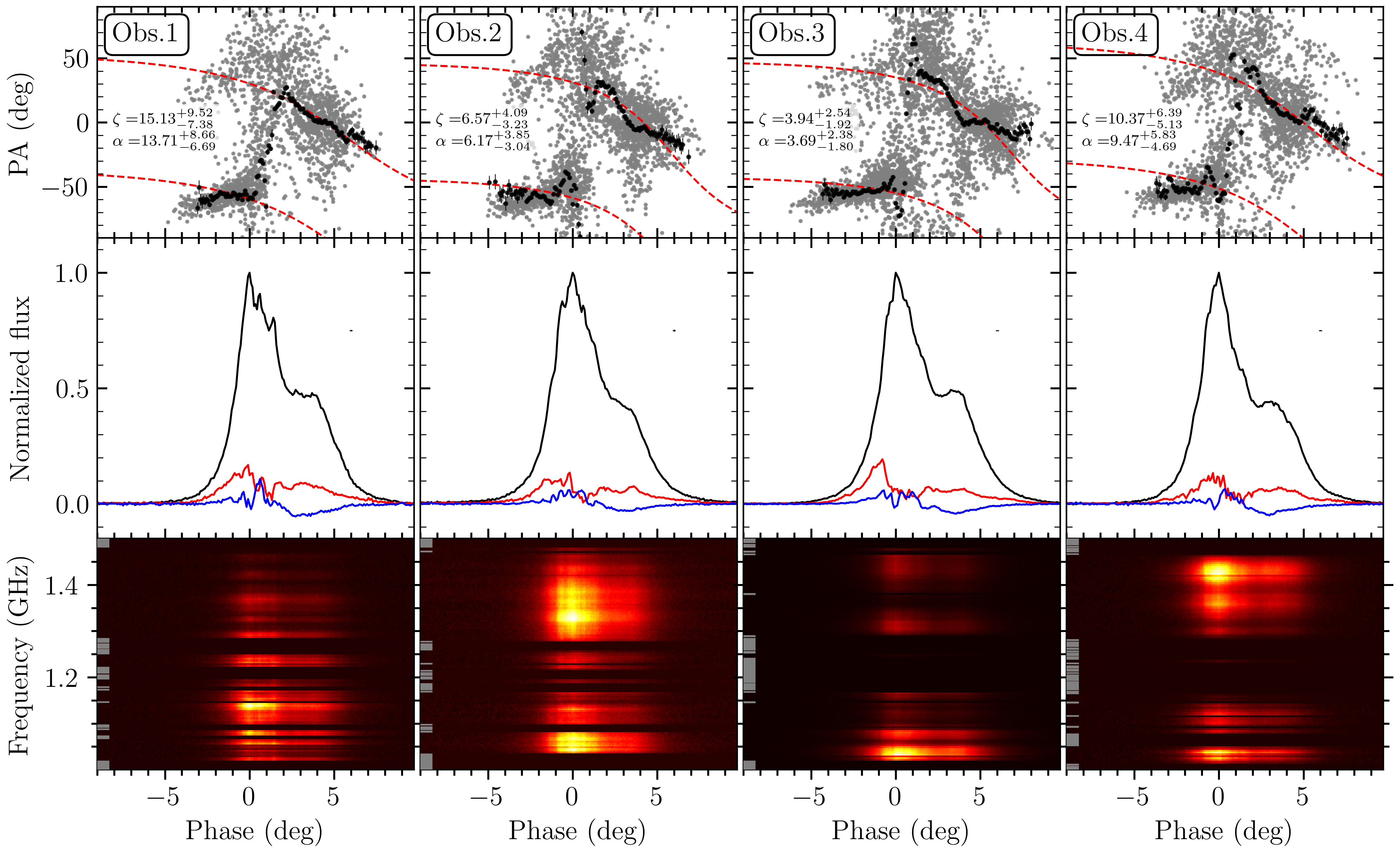}
    \caption{Polarization and integrated pulse profiles for each observation. The pulse peak is aligned at phase $0~{\rm deg}$. Top panel: Polarization position angle (PA) as a function of rotational phase. Black error bars and gray dots represent the PA of the integrated profile and single pulses, respectively. The red dashed lines indicate the RVM fit. Middle panel: Total intensity (black), linear polarization (red), and circular polarization (blue). The bin size and the root-mean-square of the off-pulse region are indicated by the error bar in the upper right region. Bottom panel: Dynamic spectrum. The gray segments on the left indicate frequency channels that were masked due to RFI. \label{fig:profile}}
\end{figure*}

\begin{figure*}
    \centering
    \includegraphics[width=\linewidth]{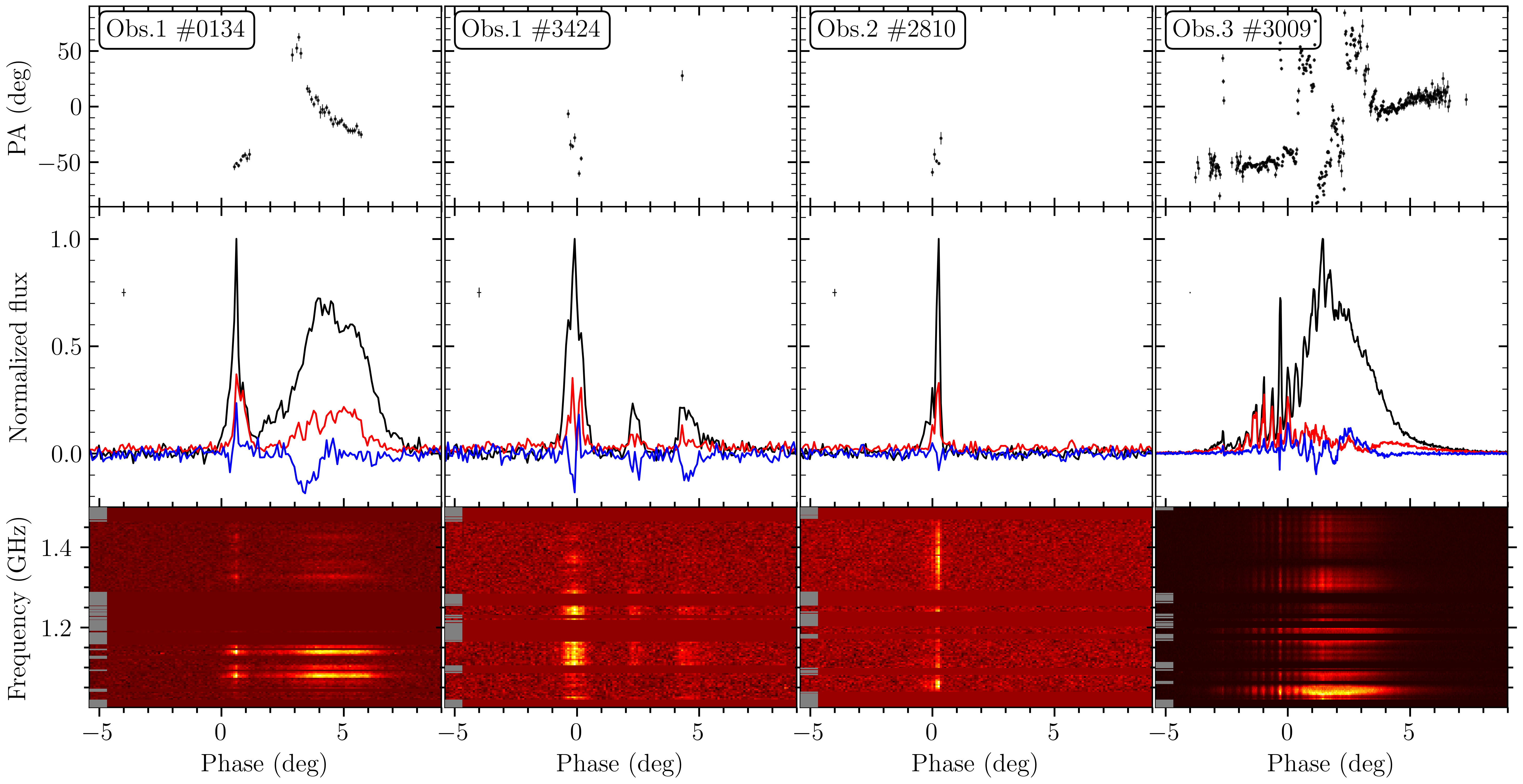}
    \caption{Similar to \autoref{fig:profile}, but for single pulses from RRAT~J2325--0530. The pulse profile has been aligned such that the peak of the pseudo-integrated pulse profile corresponds to the phase of $0~{\rm deg}$. The rightmost panel is analyzed with 16384 phase bins to resolve micro-structure, while the other pulses are analyzed using 4096 bins. \label{fig:SP}}
\end{figure*}

\begin{figure*}[htb]
    \centering
    \includegraphics[width=\linewidth]{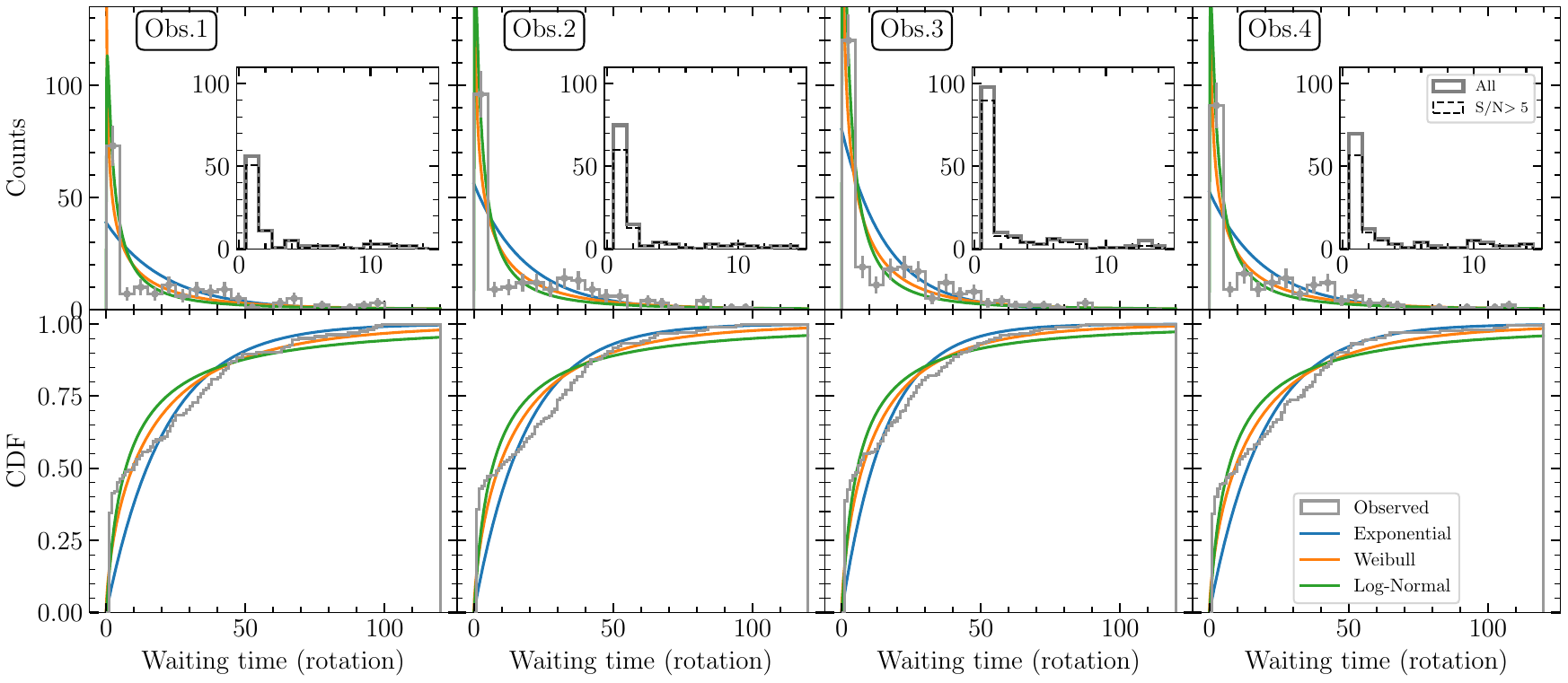}
    \caption{Waiting-time distributions for Obs.1 to 4 (left to right). Gray histograms (with a bin size of 5 rotations) show the observed waiting times, with Poisson counting errors indicated. Insets display a zoomed-in view of the range 1--15 rotations using a finer binning of 1 rotation, where the distribution for pulses with ${\rm S/N}>5$ is overplotted as a dashed-line histogram. Maximum likelihood fits using exponential, Weibull, and log-normal models are overlaid in blue, orange, and green, respectively. The top panels show the binned count distributions, while the bottom panels display the corresponding cumulative distribution functions (CDFs).
    \label{fig:wt}}
\end{figure*}

\section{Observations and data reduction}\label{sec:obs}

We observed RRAT~J2325--0530 with FAST in ``tracking'' or ``swiftcalibration'' modes, using the central beam of the $L$-band 19-beam receiver \citep{LiDi+2018,Jiang+2020}. Four observations were conducted under projects PT2021\_0117 (PI: Xiang-Dong Li) and PT2024\_0015 (PI: Shi-Jie Gao): a 50-min session in 2021, a 57.75-min session in 2024\footnote{About 1 min of this observation was affected by strong broadband radio frequency interference (RFI).}, and two sessions in 2025 lasting 57.75 min and 58.7 min, respectively. The UTC dates and MJD of each observation are listed in \autoref{tab:obs}. The observations were made at a center frequency of 1.25~GHz, covering a 500~MHz bandwidth with $\lesssim 10\%$ edge flagging \citep{LiDi+2018,Jiang+2020}. Data were recorded with $49.152~{\mu\rm s}$ time resolution across 4096 frequency channels. A noise diode was periodically injected for polarization calibration prior to each on-source observation, with the telescope pointed to an off-source position.

Pulsar ephemeris of RRAT~J2325--0530 was obtained from the ATNF pulsar catalog \citep{Manchester+2005}. The raw search-mode PSRFITS \citep{psrfits} data were processed with \texttt{DSPSR} package \citep{vanStraten+2011} into single-pulse archives with 4096 bins per rotation ($0.221~{\rm ms}$ resolution) and 1024 frequency channels, using incoherent dedispersion at ${\rm DM}=14.966~{\rm pc~cm^{-3}}$ \citep{Meyers+2019}. The \texttt{paz/pazi}, \texttt{pac}, \texttt{pav}, \texttt{rmfit} and \texttt{pam} tools from \texttt{PSRCHIVE} software suite \citep{psrfits} were used for pulsar data reduction.

We use \texttt{paz} and \texttt{clfd} tool \citep{Morello+2019} to eliminate radio frequency interference (RFI). We employed \texttt{pav} to generate summed pulse profiles and dynamic spectra for every rotation, facilitating visual inspection. To enhance sensitivity during this process, we averaged (scrunched) phase bins and frequency channels by factors of 2--8, depending on the noise level. A pulse was considered ``detection'' if a peak was present in the single-pulse profile and corresponded to a signal in the dynamic spectrum. For each detected pulse, the peak signal-to-noise ratio S/N=$I_{\rm max}/\sigma_{\rm off}$ was measured using 1024 phase bins where $I_{\rm max}$ is the maximum within the on-pulse region and $\sigma_{\rm off}$ is the root-mean-square of the off-pulse region. The corresponding peak flux density was estimated from the radiometer equation \citep[e.g.,][]{McLaughlin+2003}:
\begin{equation}
    S_{\rm peak}={\rm S/N} \times \frac{T_{\rm sys}}{G\sqrt{N_{\rm p}\Delta\nu W}},
\end{equation}
where $T_{\rm sys}=20~{\rm K}$ is the system temperature, $G=16~{\rm K~Jy^{-1}}$ is the telescope gain \citep{Jiang+2020}, $N_{\rm p}=2$ is the number of summed polarizations, $\Delta\nu$ is the effective bandwidth (taken as 75\% of the nominal 500 MHz after RFI excision), and $W=P/1024\simeq 0.85~{\rm ms}$ is the phase-bin width. The lowest S/N measured among all four observations is $\simeq 2$, corresponding to a minimum detectable peak flux density of $\sim 1.6~{\rm mJy}$, which is about two orders of magnitude deeper than the sensitivity reported for Parkes observations \citep{Meyers+2019}. We identified 4, 2, and 4 single pulses with ${\rm S/N}<3$ in Obs.2, Obs.3 and Obs.4, respectively. These pulses are likely underestimated in S/N, as they occupy only part of the full band, which reduces their measured significance. We note that this definition of a ``detection'' involves some subjectivity, and that very low S/N pulses may occasionally be misidentified. Nevertheless, we show later that applying more conservative S/N cuts does not affect our main conclusions.

For the identified single pulses, \texttt{pazi} was used to further manually eliminate narrowband and impulsive RFI. The number $N_{\rm SP}$ of identified single pulses and corresponding pulse rates are listed in \autoref{tab:obs}. The pulse rates are in the range of $196.8-263.9~{\rm hr^{-1}}$, which is $4-6$ times higher than those reported with Parkes at 1.4~GHz \citep{Meyers+2019}. We measured the observed rotation measure (RM) RM$_{\rm obs}$ from integrated pulse profile (only detected single pulses) using \texttt{rmfit} and update the RM value in the archive file headers with \texttt{pam}. The ionospheric contribution ${\rm RM}_{\rm ion}$ was calculated using the UQRG global ionospheric maps with \texttt{Spinifex} tool \citep{Mevius+2025}, allowing us to derive the interstellar medium RM via ${\rm RM}_{\rm ISM}={\rm RM}_{\rm obs}-{\rm RM}_{\rm ion}$. The observed $\rm RM_{\rm ISM}$ agree with the previously reported value of $\rm RM_{\rm ISM}=3.85\pm 0.12~{\rm rad~m^{-2}}$ at 154~MHz \citep{Meyers+2019}. The derived RM$_{\rm ISM}$ for Obs.3 differs from the other observations, likely due to uncertainties or unresolved fluctuations in the ionospheric RM correction.

\section{Results and Analysis}\label{sec:res}

\subsection{Polarization and Single-pulse Diversity}

Detected single pulses from each observation were combined to produce integrated profiles and dynamic spectra shown in \autoref{fig:profile}. Profiles were aligned so that the peak intensity corresponds to a rotational phase of $0~{\rm \deg}$. Faraday rotation was corrected using the measured RM$_{\rm obs}$, referenced to infinite frequency.
The resulting total intensity ($I$, black line), linear polarization ($L$, red line), and circular polarization ($V$, blue line) are shown in the middle panel of \autoref{fig:profile}. The summed profiles exhibit slight differences across epochs, though the overall morphological trend remains consistent.  Additional pulses \citep[$\sim$1000 pulses,][]{Liu+2012} are required to construct a more stable and representative average profile. The pulse profile exhibits a knee-like structure at phase of $\sim 4~{\rm deg}$. The integrated pulse profile and polarization characteristics are consistent with those previously reported at 1.4~GHz by the Parkes radio telescope \citep{Meyers+2019}, confirming the validity of our data reduction and calibration processes. The bottom panel of \autoref{fig:profile} shows the dynamic spectrum, where the gray segments on the left indicate frequency channels that were masked due to RFI. As shown in the dynamic spectra, bright scintles appear at varying frequencies across different observations. These patterns are consistent with interstellar diffractive scintillation, as extensively analyzed by \cite{Meyers+2019}, who demonstrated substantial modulation of intensity across the observing bandwidth at 1.4~GHz.

The polarization position angle (PA) is shown in the top panel of \autoref{fig:profile} with black error bars representing the integrated profile and gray dots indicating single-pulse measurements, respectively. The PA does not exhibit the pronounced ``S''-shape swing predicted by the standard rotating vector model \citep[RVM,][]{Radhakrishnan+1969}, Instead, it appears to contain an orthogonal polarization jump \citep{Manchester+1975} near $\sim 2~{\rm deg}$. We obtained an RVM fit for each observation (see \autoref{sec:rvm}), shown as red dashed lines in the top panels of \autoref{fig:profile}. The two lines in each panel correspond to the two orthogonal polarization components. While the fits do not fully capture all features of the observed PA, the overall trend is reasonably reproduced. The derived geometrical parameters are a line-of-sight angle $\zeta\sim 4-15~{\deg}$ and a magnetic inclination angle $\alpha\sim 4-13~{\deg}$.

\autoref{fig:SP} presents four representative single pulses of RRAT~J2325--0530, each showing the PA (top), polarization profile (middle), and dedispersed dynamic spectrum (bottom). All profiles are aligned so that the peak of the corresponding integrated profile occurs at rotational phase $0~{\rm deg}$. From left to right, the pulses illustrate the following morphologies: double peak, partially nulling, narrow, and quasi-periodic micro-structure. The micro-structure pulse (Obs3.~\#3009) shown in \autoref{fig:SP} was analyzed with 16384 phase bins while the other pulses were analyzed using 4096 bins. We estimated the quasi-periodicity ($P_{\mu}$) using a Lomb-Scargle periodogram \citep{Lomb+1976,Scargle+1982} applied to the detrended pulse profile, where the trend was removed via a two-Gaussian fit (see \autoref{sec:micro} for details). This yields $P_{\mu}=0.839^{+0.025}_{-0.034}~{\rm ms}$, which is consistent with the empirical prediction $P_{\mu} \simeq 0.94(P/{\rm 1~s})^{0.97}~{\rm ms}\simeq 0.82~{\rm ms}$ \citep[e.g.,][]{Kramer+2024}. In \autoref{sec:micro}, \autoref{fig:microSPs} presents six additional single pulses exhibiting quasi-periodic microstructures. RRAT~J2325--0530 thus becomes the fourth known RRAT to exhibit micro-structure emission, following RRATs~J1918--0449 \citep{Chen+2022}, J0139+3336 \citep{Dang+2024}, and J1913+1330 \citep{Zhong+2024,Tang+2025}.

\subsection{Waiting-time Distribution and Statistical Modeling}

\begin{table*}[htb]
    \centering
    \caption{Maximum likelihood estimation results for the waiting-time distributions of RRAT~J2325--0530. Fits are performed for exponential, Weibull, and log-normal models on both ungrouped pulses (upper block) and grouped consecutive pulses (on-window, lower block). For each model, the best-fit parameters (with $1\sigma$ standard errors from the inverse Hessian of the negative log‑likelihood at the maximum likelihood estimate), $p$-values from K-S tests and the relative AIC ($\Delta$AIC) with the minimum $\Delta$AIC rescaled to zero are presented.\label{tab:fit}}
    \begin{tabular}{c|ccc|cccc|cccc}
    \hline
    \multirow{2}{*}{Obs.}& \multicolumn{3}{c|}{Exponential} & \multicolumn{4}{c|}{Weibull} & \multicolumn{4}{c}{Log-normal}\\
    \cline{2-12}
    &$\hat \lambda~(\rm hr^{-1})$& $p$-value& $\Delta$AIC& $\hat \lambda~(\rm hr^{-1})$ & $\hat k$ & $p$-value & $\Delta$AIC & $\hat \mu$ & $\hat \sigma$ & $p$-value & $\Delta$AIC\\
    \hline
\multicolumn{12}{c}{Original single pulses}\\
\hline
1 & $197\pm15$ & $10^{-15}$ & -- & $256\pm31$ & $0.68\pm0.04$ & $10^{-6}$ & -- & $1.95\pm0.13$ & $1.68\pm0.09$ & $10^{-7}$ & -- \\
2 & $221\pm15$ & $10^{-20}$ & -- & $277\pm29$ & $0.70\pm0.04$ & $10^{-8}$ & -- & $1.89\pm0.11$ & $1.65\pm0.08$ & $10^{-10}$ & -- \\
3 & $263\pm17$ & $10^{-24}$ & -- & $331\pm31$ & $0.70\pm0.03$ & $10^{-12}$ & -- & $1.73\pm0.10$ & $1.59\pm0.07$ & $10^{-14}$ & -- \\
4 & $211\pm15$ & $10^{-16}$ & -- & $267\pm28$ & $0.70\pm0.04$ & $10^{-7}$ & -- & $1.93\pm0.11$ & $1.64\pm0.08$ & $10^{-8}$ & -- \\
\hline
\multicolumn{12}{c}{Consecutive pulses after grouping (on-window)}\\
\hline
1 & $130\pm13$ & 0.24 & 1.41 & $123\pm11$ & $1.16\pm0.09$ & 0.68 & 0.00 & $3.01\pm0.10$ & $1.08\pm0.07$ & 0.05 & 12.70\\
2 & $142\pm12$ & 0.01 & 9.65 & $132\pm9$ & $1.29\pm0.09$ & 0.36 & 0.00 & $2.97\pm0.09$ & $1.05\pm0.06$ & 0.01 & 30.14\\
3 & $161\pm13$ & 0.01 & 11.53 & $149\pm10$ & $1.28\pm0.08$ & 0.33 & 0.00 & $2.87\pm0.08$ & $0.98\pm0.06$ & 0.02 & 20.03\\
4 & $139\pm12$ & 0.10 & 3.09 & $132\pm10$ & $1.17\pm0.08$ & 0.61 & 0.00 & $2.95\pm0.09$ & $1.06\pm0.06$ & 0.02 & 16.71\\
    \hline
    \end{tabular}
\end{table*}

Waiting-time distributions for known RRATs are often modeled by exponential functions \citep[e.g.,][]{Shapiro-Albert+2018,Chen+2022,Xie+2022,Hsu+2023,Zhong+2024}. If single-pulse emission follows a Poisson process, the waiting-time ($\Delta t$) distribution is expected to be exponential, with a probability density function (PDF) given by
\begin{equation}
f_{\rm EXP}(\Delta t;\lambda) = \lambda\mathrm e^{-\lambda \Delta t},
\end{equation}
where $\lambda$ is the event rate. A more general model is the Weibull distribution \citep{Weibull+1951}, whose PDF is given by
\begin{equation}
    f_{\rm WB}(\Delta t;\lambda,k)=\lambda k(\lambda \Delta t)^{k-1}\mathrm e^{-(\lambda \Delta t)^{k}},
\end{equation}
where $k$ is the shape parameter and $\lambda$ is the rate parameter. The Weibull distribution has been applied to waiting times for repeating FRBs \citep{Oppermann+2018} and RRATs \citep{Hsu+2023}. When $k<1$, clustering is present with lower $k$ values corresponding to more clustering. When $k=1$, the Weibull distribution reduces to the exponential distribution, characteristic of Poisson process with memoryless behavior. For $k>1$, bursts become more probable with time since the last event, often implying a periodic or relaxation-driven mechanism. We also consider the log-normal distribution which provides a good description for FRBs. Its PDF is given by
\begin{equation}
    f_{\rm LN}(\Delta t/P;\mu,\sigma)=\frac{1}{(\Delta t/P)\sigma\sqrt{2\pi}} \mathrm e^{-\frac{\left[\ln (\Delta t/P)-\mu\right]^2}{2\sigma^2}},
\end{equation}
where $\mu$ and $\sigma$ are the mean and standard deviation of $\ln (\Delta t/P)$, respectively. Here, $\Delta t$ is normalized by the spin period $P$, so that $\Delta t/P$ is dimensionless, as required for the logarithm.

The top panels of \autoref{fig:wt} display the RRAT~J2325--0530's waiting‐time distributions (gray histograms) for each observation, using a bin width of 5 rotations. A striking feature emerges: the count of pulses separated by just one rotation (consecutive bursts) is unusually high. This is highlighted in the insets, where we zoom in with a finer 1‑rotation bin size. The insets also show the distribution for pulses with ${\rm S/N}>5$. The one-rotation peak remains clearly significant, indicating that the observed clustering is not strongly dependent on faint single pulses. The bottom panels show the corresponding cumulative distribution functions (CDFs) using 1‑rotation bins. In each case, there is a pronounced jump at one rotation (CDF$\sim 35\%$), confirming that a substantial fraction of bursts occur in immediate succession. In a Poissonian process, waiting times follow an exponential distribution that peaks at the minimum interval, yet the observed frequency of one‑rotation intervals for RRAT~J2325--0530 greatly exceeds the expectation. This clear excess reveals a departure from pure Poisson behavior, instead indicating a strong tendency for pulses to cluster tightly in time.

For each observation of RRAT~J2325--0530, we estimated the parameters of the waiting-time distributions using maximum likelihood estimation. The estimated parameters are summarized in Table~\ref{tab:fit}. In \autoref{fig:wt}, the blue, orange, and green curves represent the fitted distributions: exponential, Weibull, and log-normal, respectively. These are shown in both the counts (top three panels) and the cumulative distribution function (bottom three panels). As seen in \autoref{fig:wt}, none of the tested distributions provide a satisfactory fit to the observed data.

To quantitatively assess the goodness of fit, we apply the Kolmogorov-Smirnov (K-S) test. The null hypothesis ($H_0$) assumes that the observed waiting times follow the specified distribution, while the alternative hypothesis ($H_1$) assumes they do not. A high $p$-value indicates insufficient evidence to reject $H_0$, suggesting that the model may plausibly describe the data. Conversely, a low $p$-value (typically less than 0.05) leads to the rejection of $H_0$, indicating that the model does not adequately describe the data. For all combinations of datasets and distributions tested, the K-S test yields $p$-values much smaller than 0.05. This result indicates that the waiting-time distribution of RRAT~J23250--0530 can not be described by any of the three tested distributions.

\subsection{Interpretation of Burst Clustering}

\begin{figure*}
    \centering
    \includegraphics[width=\linewidth]{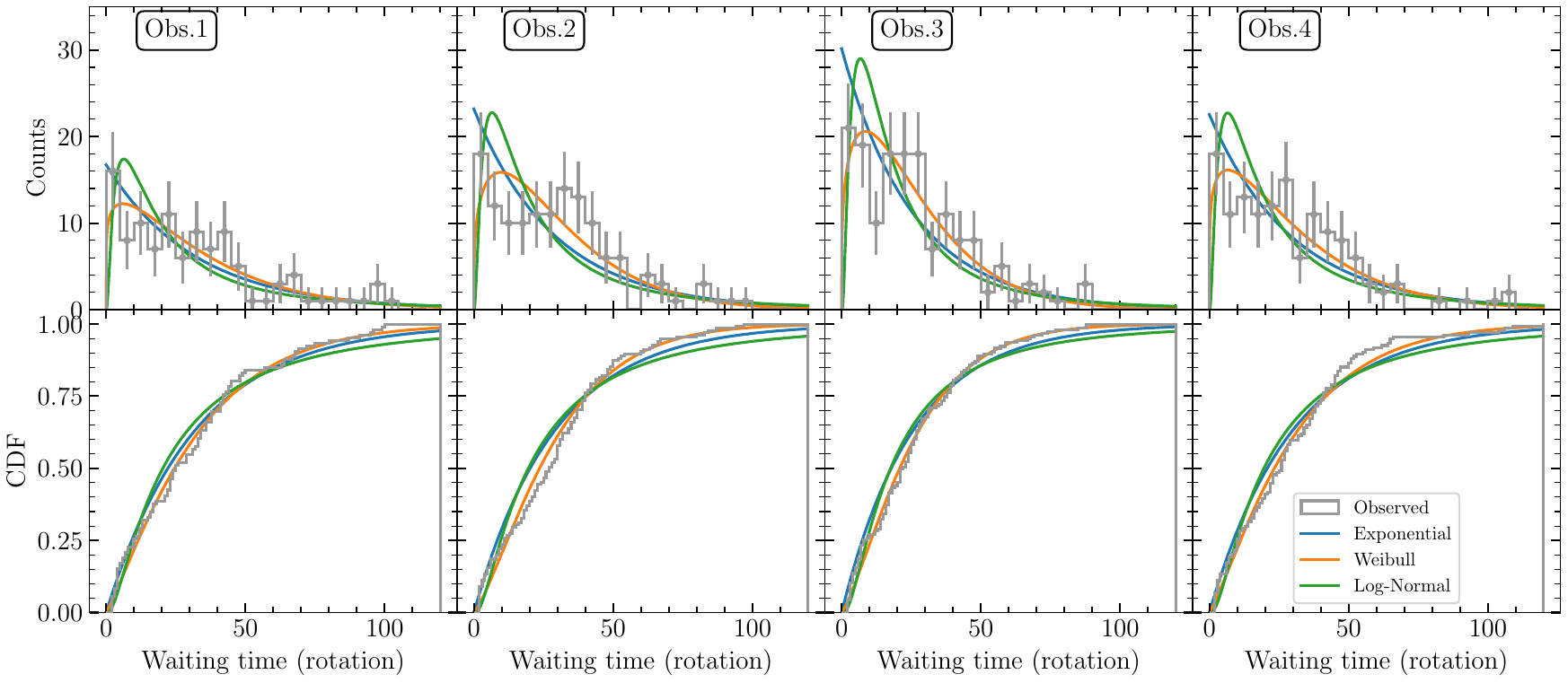}
    \caption{Similar to \autoref{fig:wt}, but with consecutive bursts combined and treated as a single burst, representing the waiting‑time distributions within the on‑window intervals.}
    \label{fig:wt2}
\end{figure*}

\begin{figure*}
    \centering
    \includegraphics[width=\linewidth]{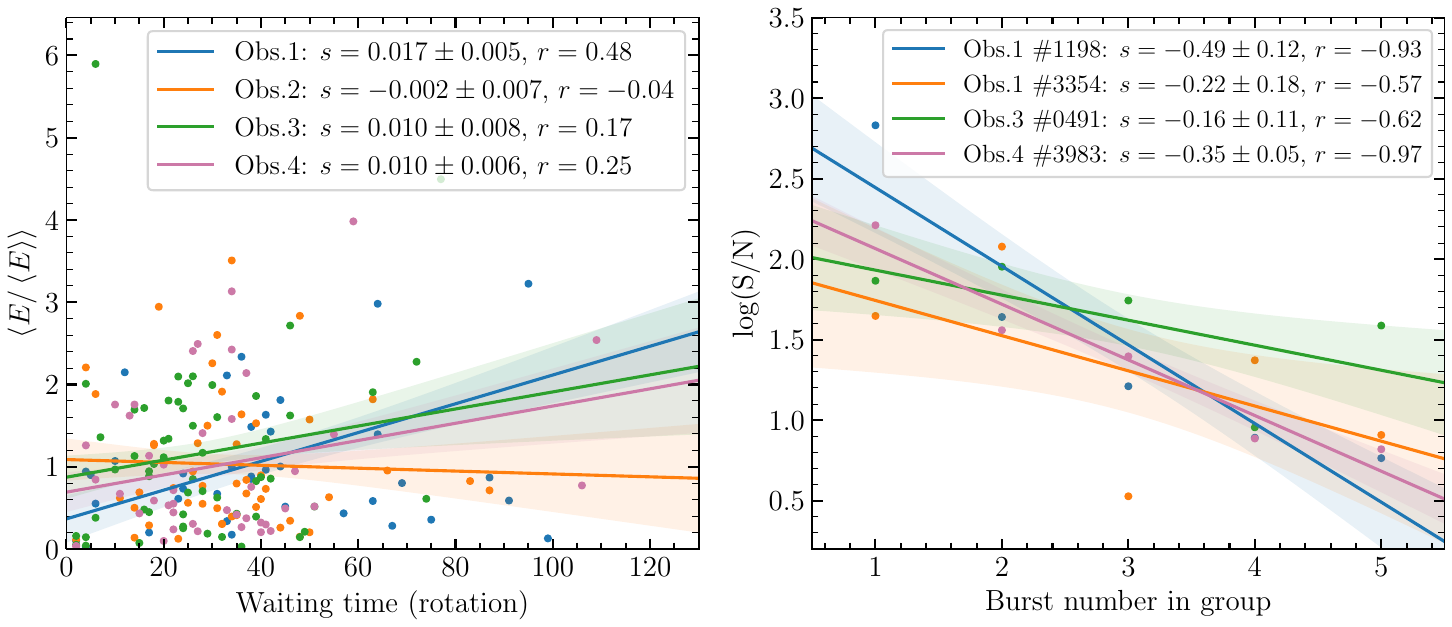}
    \caption{Left: Correlation between average normalized single-pulse energy ($\left<E/\left<E\right>\right>$) and waiting time for burst groups of RRAT~J2325--0530 across four FAST observations. Solid lines show linear fits with shaded regions indicating $1\sigma$ uncertainties. Colored points correspond to the each observation, as indicated in the legend along with the fitted slope $s$ and the Pearson correlation coefficient $r$. Right: Peak signal-to-noise ratio $\log ({\rm S/N})$ within five-pulse groups (see \autoref{fig:5}). Solid lines and shaded regions show linear fits and their $1\sigma$ uncertainties. Colored points match the corresponding burst groups. The legend shows the slope $s$ and Pearson correlation coefficient $r$.\label{fig:E}}
\end{figure*}

To interpret the atypical waiting-time behavior observed in RRAT~J23250--0530, we propose a simple emission model in which consecutive pulses represent a single burst event, modulated by the pulsar's rotation. In this scenario, the emission region is active for a duration (on-window) slightly longer than the rotation period, but the beam is only visible to the observer once per rotation. Thus, each time the beam sweeps across our line of sight, we detect a single pulse, resulting in consecutive single-pulse detections. This picture resembles the RRAT model proposed by \citet{Burke-Spolaor+2010}, in which RRATs exhibit emission windows shorter than one rotation period. Our variation assumes that the on-window slightly exceeds one rotation period. 

To test this model, we group consecutive bursts (separated by exactly one rotation) as a single burst spanning multiple rotations. The multiplicity distribution of these grouped bursts, i.e., the number of burst sequences containing 1, 2, 3, 4 or 5 consecutive pulses, is listed in \autoref{tab:obs}. The longest sequence contains five consecutive bursts observed in Obs.1, 3 and 4, and no group contains six or more. The aligned pulse profiles of four five-pulse groups are shown in \autoref{fig:5}. We then recalculate the waiting-time distributions within the on‑window intervals, as shown in \autoref{fig:wt2}.

The on‑window waiting-time distributions superficially resembles exponential decay. To determine which model best describes the data, we also performed maximum likelihood estimations for three distributions: exponential, Weibull, and log-normal. The resulting best-fit parameters are listed in \autoref{tab:fit}, and the model fits are shown in both the counts and CDF panels of \autoref{fig:wt2}. K-S tests were used to assess the goodness of fit, with $p$-values reported in \autoref{tab:fit}. The exponential model yields a $p$-value greater than 0.05 for Obs.1, suggesting null hypothesis cannot be rejected. However, for the other three observations, the exponential model is rejected. In contrast, the Weibull distribution fits all four observations successfully, with $p$-value $>0.05$ in every case. The log-normal model is rejected for all observations. This is consistent with the visual impression from \autoref{fig:wt2}, which shows that the Weibull model (orange curves) best matches the observed distributions, especially for Obs.3. We note a modest excess in the distribution at $\sim 25-50$ rotations in Obs.2 and Obs.4. Similar features have previously reported in RRATs~J1819--1458 and J1317--5759 \citep{Shapiro-Albert+2018}. This excess causes a slight deviation from the ideal Weibull distribution, particularly in Obs.4, and may reflect additional timescale of pulse clustering beyond the simple Weibull model.

Because $p$-values of K-S tests are not directly comparable across models, we also compute the Akaike Information Criterion (AIC) to evaluate relative model performance, i.e., ${\rm AIC}=2n-2\ln \mathcal{L}$, where $n$ is the number of model parameters and $\mathcal{L}$ is the maximized likelihood function. The model with the lowest AIC is preferred. We rescale the minimum AIC value to zero and use the relative difference $\Delta$AIC for comparison. In all four observations, the Weibull distribution has the lowest AIC, strongly favoring it over the exponential and log-normal models. The $\Delta$AIC values for the alternative models range from 1.4 to 30, indicating significantly weaker support.

The best-fit Weibull shape parameters are $k \gtrsim 1$, indicating that the burst intervals exhibit some deviations from a purely random process and that the likelihood of burst occurrence may increase with time since the previous pulse. This behavior may be consistent with an energy store-release mechanism \citep[e.g.,][]{Lyutikov+2002}. To examine whether the burst energy within a group correlates with the on‑window waiting time, we performed linear regression between the average normalized pulse energy $\left<E/\left<E\right>\right>$ within a pulse group and the waiting time of single-pulse groups.
Here, $E$ denotes the fluence (energy) of an individual pulse, and $\left<E\right>$ represents the mean single-pulse energy of each observation. The results shown in \autoref{fig:E} reveal no strong correlation, with Pearson correlation coefficients $r$ ranging from $-0.04$ to $0.48$, varying across the four observations. These results are consistent with previous studies of RRATs and nulling pulsars, which also report little or no correlation between burst energy and waiting time \citep{Gajjar+2012,Shapiro-Albert+2018,Hsu+2023}.

We show the peak signal-to-noise ratio S/N for each pulse within groups containing five pulses using 4096 phase bins (as illustrated in \autoref{fig:5}). As seen in \autoref{fig:5}, S/N tends to decrease over time within each burst group, particularly for Obs.1~\#1198--1202  and Obs.4~\#3983--3987. To quantify this trend, we performed linear regressions between $\log({\rm S/N})$ and the time since the first burst in each group. The results, shown in the left panel of \autoref{fig:E}, reveal moderate/strong negative correlations, with Pearson correlation coefficients $r$ ranging from $-0.57$ to $-0.97$. Notably, the coefficients $r$ for Obs.1~\#1198--1202  and Obs.4~\#3983--3987 are $-0.93$ and $-0.97$, respectively, indicating strong negative correlations. This suggests that the burst intensity within an on-window decays approximately exponentially.

\begin{figure*}
    \centering
    \includegraphics[width=\linewidth]{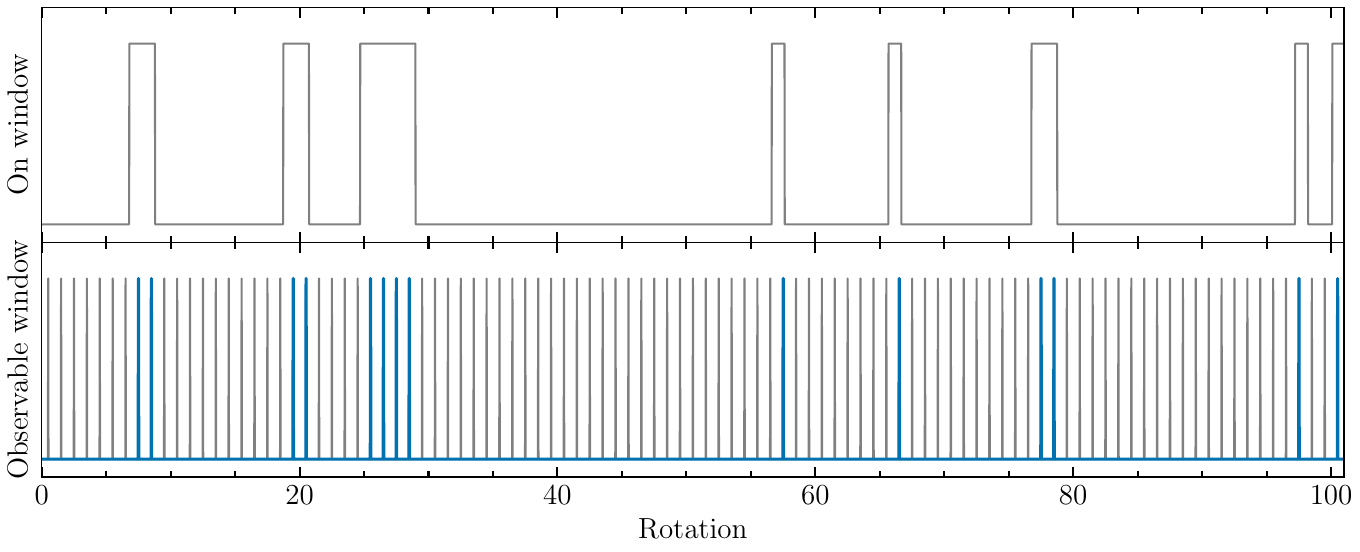}
    \caption{Simulated burst behavior of RRAT~J2325--0530. The top panel shows burst on-windows drawn from a Weibull distribution. The bottom panel shows observable windows modulated by the pulsar's spin. Blue segments indicate detectable single pulses where the two windows overlap.\label{fig:sketch}}
\end{figure*}

\begin{figure*}
    \centering
    \includegraphics[width=\linewidth]{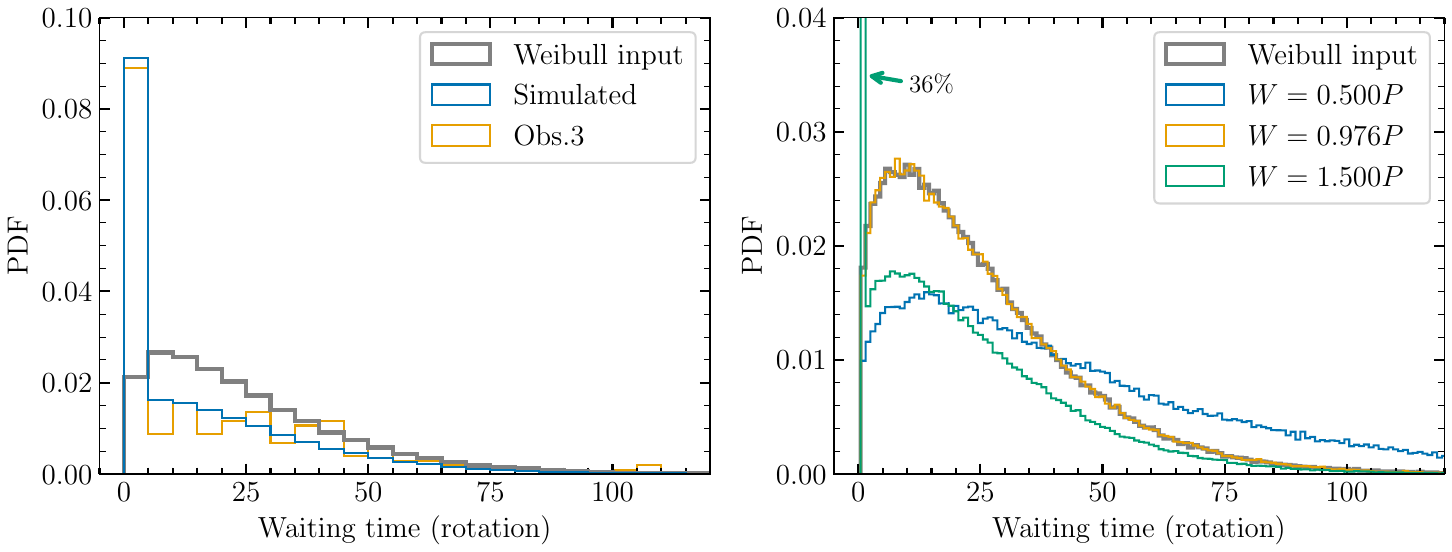}
    \caption{Monte Carlo simulations of the waiting time distributions for RRAT~J2325--0530. Left: Comparison between the input Weibull distribution (gray), simulated distribution for Obs.3 (blue), and observed distribution of Obs.3 (orange). Right: Simulations with fixed burst widths of $W=0.5P$, $W=P-w=0.976P$ and $W=1.5P$. A clear one-rotation peak appears when $W+w>P$ and 36\% of the simulated single pulses fall at one rotation (not visible due to the y-axis being truncated at 0.04).\label{fig:mc}}
\end{figure*}

\subsection{Monte Carlo Simulations}

We conducted Monte Carlo simulations to reproduce the observed waiting-time distribution, focusing on Obs.3, which has the highest burst rate and the largest number of detected bursts. In the simulation, intrinsic burst interval (on-window) were draw from a Weibull distribution with parameters $\lambda=149~{\rm hr^{-1}}$ and $k=1.28$. The on-window width $W$ were sampled based on the multiplicity statistics observed in Obs.3, using weights of 98, 27, 20, 9, and 1 for groups of 1 to 5 pulses, respectively. Each individual pulse was assigned a fixed on-pulse phase width $w=0.0248P$, corresponding to the observed 10\% peak width of the integrated profile. The resulting pulses were then modulated by the pulsar's rotation.

\autoref{fig:sketch} illustrates a schematic simulation over 100 rotations. The top panel shows the intrinsic burst window (the so-called on-window) drawn from a Weibull distribution. The bottom panel depicts the spin-modulated observable window.  Observable single pulses occur when the emission beam overlaps our line of sight, highlighted in blue. The left panel of \autoref{fig:mc} compares the simulated (blue) and observed (orange) waiting-time distributions for Obs.3 with the input Weibull distribution for on-window (gray). Minor discrepancies arise from simplifications, such as using a fixed on-pulse width $w$ and excluding sub-period on-window ($W<P$), but overall the match supports the validity of our emission model.

We also simulated three fix-width burst models with $W=0.5P$, $W=P-w=0.976P$ and $W=1.5P$, shown in the right panel of \autoref{fig:mc}. For $W=0.5P$, where $W+w<P$, the probability of detecting a pulse within one rotation is $(W+w)/P$, leading to a flatter distribution. When $W=P-w=0.976P$, the waiting-time distribution matches the input Weibull distribution. In contrast, for $W=1.5P$, where $W+w>P$, a strong one-rotation peak emerges, with approximately 36\% of single pulses separated by exactly one rotation. These simulations suggest that the average burst on‑window of RRAT~J2325--0530 likely exceeds its spin period, explaining the observed one‑rotation excess in the waiting‑time distribution and reflecting characteristic magnetospheric on‑timescales.

\section{Discussion and Summary}\label{sec:sum}

\citet{Burke-Spolaor+2010} proposed that RRATs may represent an evolutionary phase in the life of a pulsar. In this framework, RRATs are interpreted as extreme nulling pulsars whose emission on-windows are shorter than a single rotation period. At such a stage, the radio emission mechanism becomes active only briefly and intermittently, producing infrequent single pulses. The waiting-time distribution we observed for RRAT~J2325--0530 appears broadly consistent with this picture, particularly the tendency for bursts to cluster within short intervals. 
However, some caveats remain. According to the spin period of $P\simeq0.869~{\rm s}$, the period derivation of $\dot P\simeq 1.03\times10^{-15}~{\rm s~s^{-1}}$ and the inferred surface dipolar magnetic field $B\simeq 9.57\times10^{11}~{\rm G}$, RRAT~J2325--0530 lies above the classical pulsar death line \citep{Ruderman+1975,Chen+1993}, though it may approach the upper boundary of the so-called death valley seen in some models \citep{Beskin+2022,Gencali+2021}. This challenges a straightforward interpretation in terms of magnetic field decay. Furthermore, many pulsars with high nulling fractions are found close to the death line without exhibiting RRAT-like behavior, and conversely, several RRATs (including RRAT~J2325--0530) are located far from it \citep{Abhishek+2022}. These discrepancies suggest that the transition to RRAT-like emission may not be governed solely by age or surface dipolar magnetic field strength, but instead may reflect more complex magnetospheric dynamics.

Observations of RRAT~J1913+1330 with FAST have revealed pulse clusters (referred to as ``active phases'') lasting between 68 and 495 rotations \citep{Lu+2019,Zhang+2024,Zhong+2024}. Due to limited observational length, the waiting-time distribution between active phases could not be determined. As shown in \autoref{fig:mc}, if RRAT~J1913+1330 shared the same burst behavior as RRAT~J2325--0530, its waiting-time distribution would be expected to show a prominent one-rotation peak. However, the observed distribution retains an exponential form \citep{Zhong+2024}. The sporadic burst activity of RRATJ1913+1330 has been attributed to pulse-to-pulse variability \citep{Zhang+2024}, which naturally explains the absence of a one-rotation peak. In this context, RRAT~J1913+1330 is unlikely to be a sibling of RRAT~J2325--0530. This discrepancy raises an important consideration regarding the sensitivity limits of observations. While the higher modulation index of RRAT~J1913+1330 may obscure a one-rotation peak, it's also possible that such a feature is simply below the detection threshold of FAST. More sensitive future observations may reveal clustering in J1913+1330, so the lack of a one-rotation peak here does not rule out its existence at lower flux density levels.

Our observations with FAST have revealed a significant increase in pulse rates compared to those reported from less sensitive telescopes. This trend is consistent with findings from other FAST observations of RRATs \citep[e.g.,][]{Lu+2019,Xie+2022,Chen+2022,Zhong+2024,Dang+2024,Zhang+2024}, indicating that burst rate measurements are highly sensitive to the sensitivity limitations of the observing instruments. This suggests that many RRATs may have underestimated pulse rates due to detection thresholds of previous surveys. While waiting time distributions are not strongly correlated with pulse energy, the underdetection of faint pulses (such as pulses fainter than \#3356 in Obs.1, see \autoref{fig:5}) can result in misleading pulse cluster behavior. Additionally, less sensitive observations lead to smaller sample sizes and reduced statistical significance in waiting time analyses. In light of this, we suggest future studies observed with simultaneous various telescopes should consider the impact of sensitivity limits when interpreting RRAT behavior.

Finally, we summarize this work as follows. We have presented four FAST observations of RRAT~J2325--0530 at 1.25~GHz. The waiting-time distribution exhibits a pronounced peak at one-rotation period, driven by uninterrupted consecutive bursts. Within five-pulse groups, the peak signal-to-noise ratio of individual pulses tends to decrease over time. After grouping consecutive bursts, the burst on-window waiting-time distribution follows a Weibull form with shape parameter $k\gtrsim 1$. This is the first observation to report such a phenomenon in RRATs. Monte Carlo simulations incorporating burst windows slightly longer than the spin period successfully reproduce the observed one-rotation peak. Although these findings may suggest that RRAT~J2325--0530 occupies a transitional stage between nulling pulsars and RRATs with shorter on-windows (as proposed by \citealt{Burke-Spolaor+2010}), caveats still remain. In particular, the transition to RRAT-like emission may not be governed solely by age or surface dipolar magnetic field strength, but could instead reflect more complex and diverse magnetospheric dynamics. Additionally, we detect quasi-periodic micro-structure in several single pulses, with a measured micro-structure period $P_{\rm \mu}$ consistent with the empirical relation $P_{\mu} \simeq 0.94(P/{\rm 1~s})^{0.97}~{\rm ms}$ \citep{Kramer+2024}.

\begin{acknowledgments}
We are grateful to the anonymous referee for carefully reading and providing insightful comments.
S.J.G. would like to thank Jin-Tao Xie for helpful discussions on polarization calibration of FAST data.
This work made use of the data from FAST (Five-hundred-meter Aperture Spherical radio Telescope, \url{https://cstr.cn/31116.02.FAST}). FAST is a Chinese national mega-science facility, operated by National Astronomical Observatories, Chinese Academy of Sciences.
S.J.G. acknowledges support from the National Natural Science Foundation of China (NSFC) under grant No.~123B2045.
X.D.L. acknowledges support from the National Key Research and Development Program of China (2021YFA0718500), the National Natural Science Foundation of China (NSFC) under grant No.~12041301, 12121003 and 12203051.
P.Z. acknowledges support from the National Natural Science Foundation of China (NSFC) under grant No.~12273010.
The computation was made by using the facilities at the High-Performance Computing Center of Collaborative Innovation Center of Advanced Microstructures (Nanjing University).
\end{acknowledgments}

\section*{Data Availability}
FAST data are open source in the FAST Data Center according to the FAST data one-year protection policy.\\

\facilities{The Five-hundred-meter Aperture Spherical radio Telescope (FAST).}

\software{\texttt{Astropy} \citep{Astropy}, \texttt{Bilby} \citep{bilby_paper}, \texttt{clfd} \citep{Morello+2019}, \texttt{DSPSR} \citep{vanStraten+2011}, \texttt{Matplotlib} \citep{Matplotlib}, \texttt{NumPy} \citep{NumPy}, \texttt{PSRCHIVE} \citep{psrfits}, \texttt{SciPy} \citep{scipy} and \texttt{Spinifex} \citep{Mevius+2025}.}

\appendix
\restartappendixnumbering
\section{RVM Fitting}\label{sec:rvm}
In the RVM, radio emission is beamed along the magnetic field lines, and the observed PA is determined by the instantaneous orientation of the magnetic field as it sweeps across the observer's line of sight \citep{Radhakrishnan+1969,Everett+2001}. The PA $\Psi$ as a function of pulse longitude $\phi$ is given by
\begin{equation}
\Psi=\Psi_0+\arctan\left[\frac{\sin \alpha\sin(\phi-\phi_0)}{\sin \zeta \cos\alpha-\cos\zeta\sin\alpha\cos(\phi-\phi_0)}\right],
\end{equation}
where $\phi_0$ is the pulse longitude at which $\Psi=\Psi_0$, $\zeta$ is the angle between the line of sight and the spin axis, $\alpha$ is the magnetic inclination angle. 
In this formulation, $\Psi$ increases clockwise on the sky, opposite to the astronomical convention (PSR/IEEE convention, \citealt{psrfits}) adopted here. To account for this, we invert the sign of the numerator in the equation.
We adopt the method of \cite{vanStraten+2012} and \cite{Desvignes+2019}, which fits the complex linear polarization (Stokes $Q$ and $U$ are treated as the real and imaginary components, respectively) predicted by the RVM. We use the Bayesian inference framework \texttt{Bilby} \citep{bilby_paper} as the sampler. This approach naturally models orthogonal polarization mode transitions as negative complex magnitudes, removing the need for manual classification \citep{vanStraten+2012}. Data points near the orthogonal polarization mode transitions were removed, and the corresponding excluded phases are listed in \autoref{tab:rvm}. Since a $180\deg$ change in pulse phase $\phi$ corresponds to viewing the opposite magnetic pole, we restrict $\phi_0$ to $[-90~\deg,~90~\deg]$ to ensure a unique determination of $\alpha$.

\begin{figure}
\plottwo{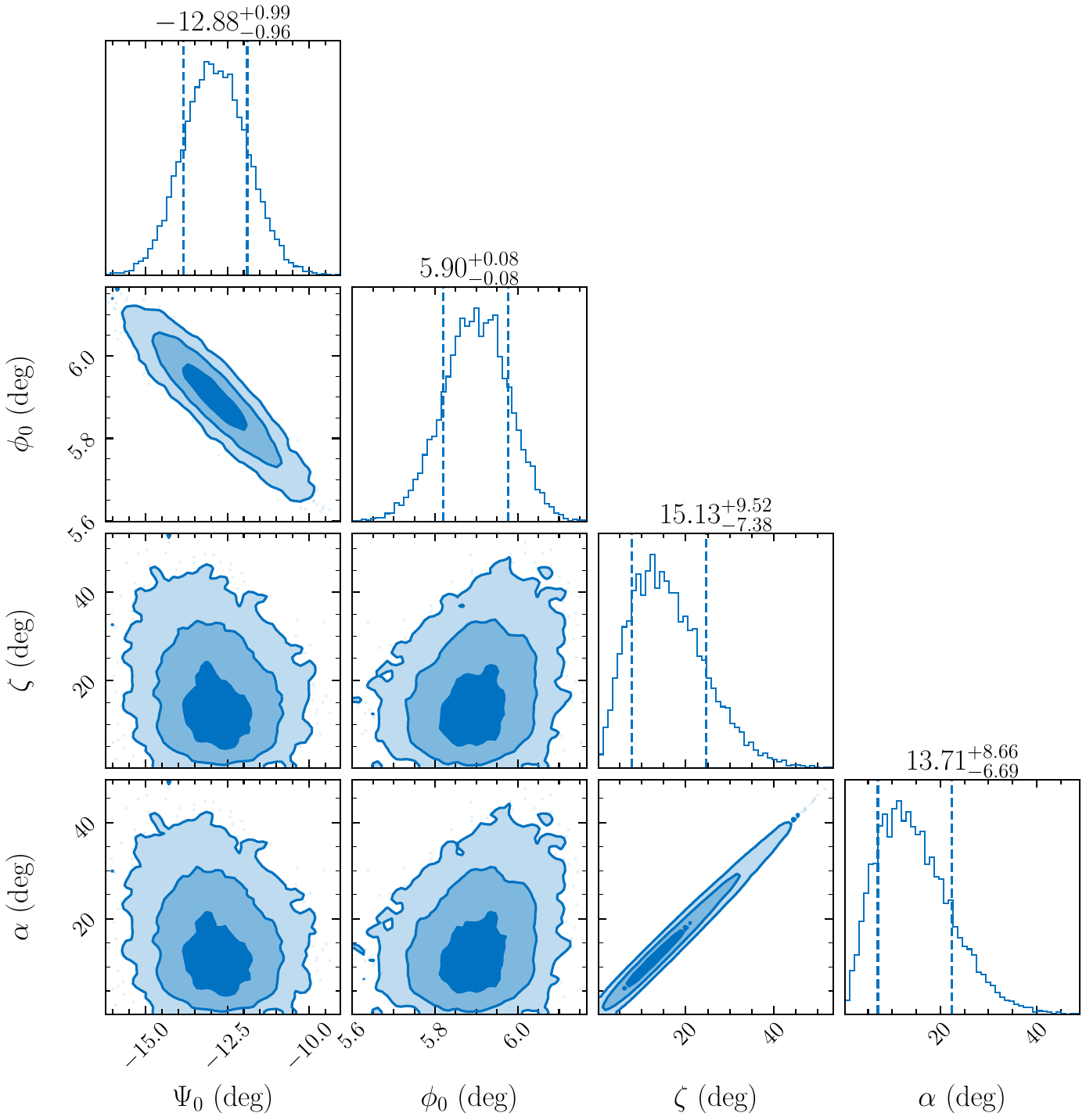}{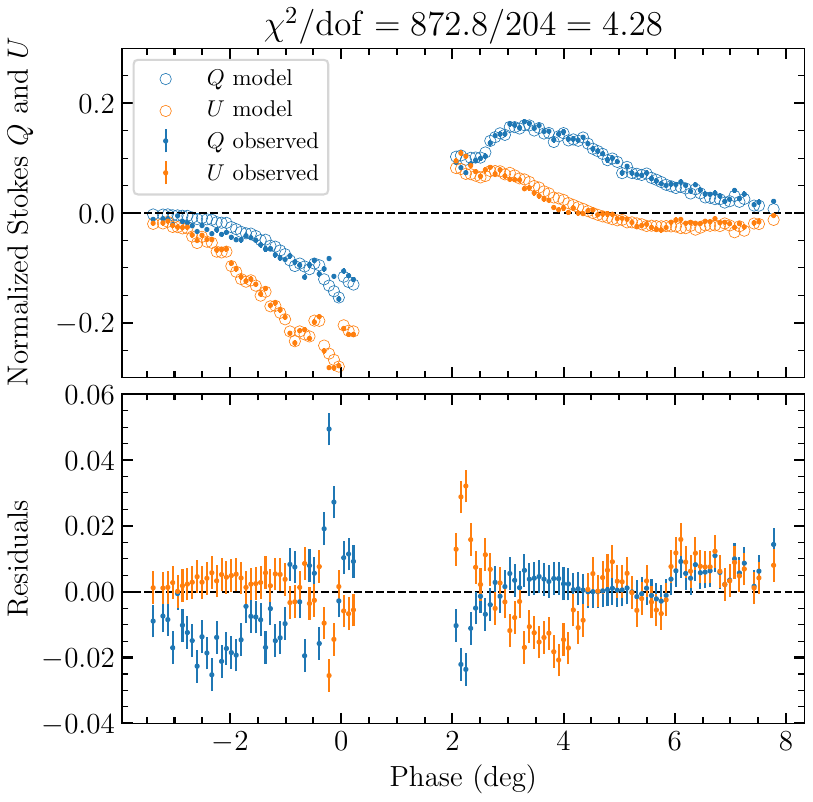}
\caption{Example RVM fit for Obs.1. Left: corner plot of the posterior distributions for the model parameters. Right: observed (points) and modeled (open circles) Stokes $Q$ and $U$ profiles, normalized by the maximum intensity, along with the residuals shown in the same units. \label{fig:rvm}}
\end{figure}

\begin{table*}
    \centering
    \caption{Best-fit RVM parameters for the four observations. The quoted uncertainties correspond to the 16\% and 84\% posterior quantiles. The reduced $\chi^2_{\rm r}$ values and the degrees of freedom (dof) are also given. The phase ranges omitted near orthogonal polarization mode transitions are given in the second column.\label{tab:rvm}}
    \begin{tabular}{ccccccccc}
    \hline
Obs.&Omitted Phase&$\Psi_0$&$\phi_0$&$\zeta$&$\alpha$&$\chi^2_{\rm r}$&dof\\
&(deg)&(deg)&(deg)&(deg)&(deg)&&\\
\hline
1&$0.18-2.02$&${-12.88}_{-0.96}^{+0.99}$&${5.90}_{-0.08}^{+0.08}$&${15.13}_{-7.38}^{+9.52}$&${13.71}_{-6.69}^{+8.66}$&4.28&204\\
2&$-0.26-0.88$&${-24.58}_{-0.50}^{+0.53}$&${5.92}_{-0.03}^{+0.03}$&${6.57}_{-3.23}^{+4.09}$&${6.17}_{-3.04}^{+3.85}$&35.19&246\\
3&$-0.18-1.49$&${-23.05}_{-0.28}^{+0.28}$&${6.83}_{-0.02}^{+0.02}$&${3.94}_{-1.92}^{+2.54}$&${3.69}_{-1.80}^{+2.38}$&48.01&260\\
4&$0.18-1.49$&${-19.20}_{-0.18}^{+0.18}$&${1.46}_{-0.01}^{+0.01}$&${5.21}_{-2.62}^{+3.21}$&${5.12}_{-2.58}^{+3.15}$&24.31&234\\
\hline 
    \end{tabular}
\end{table*}

An example fit is shown in \autoref{fig:rvm}, where the left panel presents the parameter posteriors and the right panel compares the observed and modelled normalized Stokes $Q$ and $U$ profiles. The parameters $\zeta$ and $\alpha$ are strongly correlated. Results for all four observations are listed in \autoref{tab:rvm}. The reduced chi-squared values $\chi^2_{\rm r}$ range from 4 to 48, indicating that the fits are formally poor (see also the top panels of \autoref{fig:profile}).

\section{Quasi-periodic micro-structure Analysis}\label{sec:micro}

\begin{figure*}
    \centering
    \includegraphics[width=\linewidth]{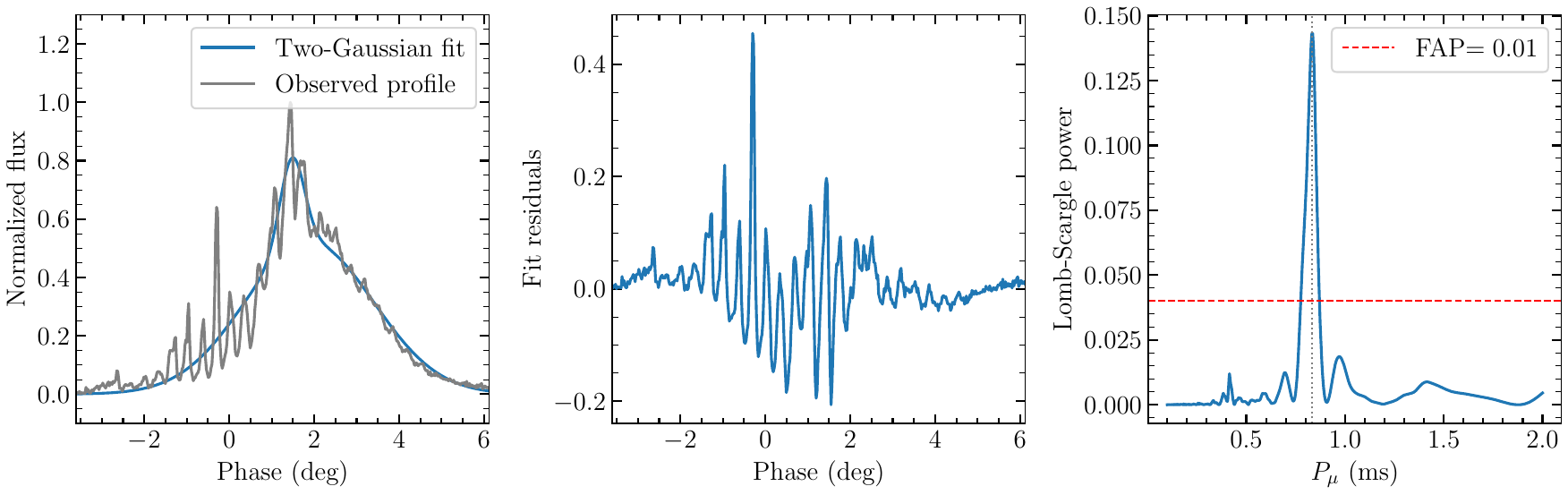}
    \caption{Quasi-periodicity analysis. Left: The integrated pulse profile (gray) is fit with a two-Gaussian model (blue). Middle: Residuals between the observed profile and the two-Gaussian fit. Right: Lomb-Scargle periodogram of the residuals. The horizontal dashed line marks the false alarm probability (FAP) threshold of 0.01, indicating that the detected peak is statistically significant.\label{fig:micro}}
\end{figure*}

\begin{figure}
    \centering
    \includegraphics[width=\linewidth]{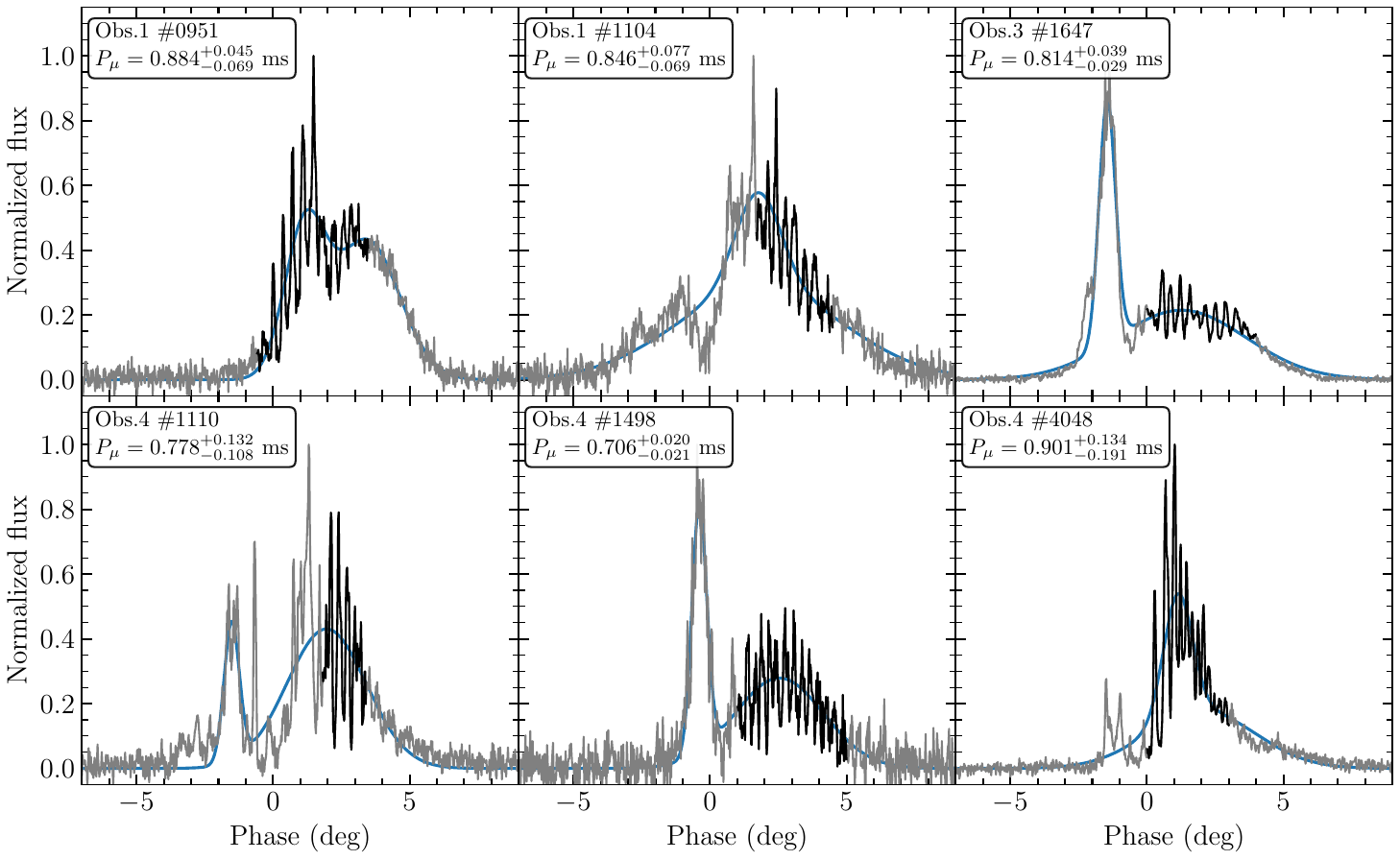}
    \caption{Examples of six additional single pulses exhibiting quasi-periodic microstructures, similar to the left panel of \autoref{fig:micro}. The black-shaded segments indicate the regions used for the periodicity analysis. The derived $P_{\mu}$ for each pulse is given in the corresponding panel.\label{fig:microSPs}}
\end{figure}

During the four FAST observations, several single pulses exhibit quasi-periodic microstructures. As an example, we analyze burst Obs.3~\#3009. We first modeled the pulse profile using a two-component Gaussian fit (left panel of \autoref{fig:micro}), which captures the main pulse morphology. The residuals from the fit (middle panel of \autoref{fig:micro}) reveal significant modulations. We applied a Lomb-Scargle periodogram \citep{Lomb+1976,Scargle+1982} to the residuals (see the right panel of \autoref{fig:micro}) and identified a significant periodicity at $P_{\mu}=0.839^{+0.025}_{-0.034}~{\rm ms}$. The uncertainty corresponds to the half-power width of the Lomb-Scargle peak, which is narrower than the phase bin width ($0.053~{\rm ms}$). The horizontal dashed line in \autoref{fig:micro} represents the false alarm probability (FAP) threshold of 0.01 ($\simeq 2.5\sigma$) and the peak corresponds a false alarm probability of $10^{-18}$, indicating that the detected peak is statistically significant.

\autoref{fig:microSPs} presents six additional single pulses exhibiting quasi-periodic microstructures. These features are confined to specific portions of the on-pulse region. Therefore, only the black-shaded segments were used for the periodicity analysis. The resulting microstructure periods $P_{\mu}$ (indicated in each panel of \autoref{fig:microSPs}) range from $0.71$ to $0.90~{\rm ms}$, consistent with model prediction of $P_{\mu} \simeq 0.94(P/{\rm 1~s})^{0.97}~{\rm ms}\simeq0.82~{\rm ms}$ \citep{Kramer+2024}. The quasi-periodic properties of Obs.4 \#4048 appear to vary over time: at the beginning, the spikes are spaced farther apart, while towards the end they become closer. This leads to a larger uncertainty in the measured period compared to the other pulses.

\section{Pulse profiles in five-pulse groups}
\begin{figure*}
    \centering
    \includegraphics[width=\linewidth]{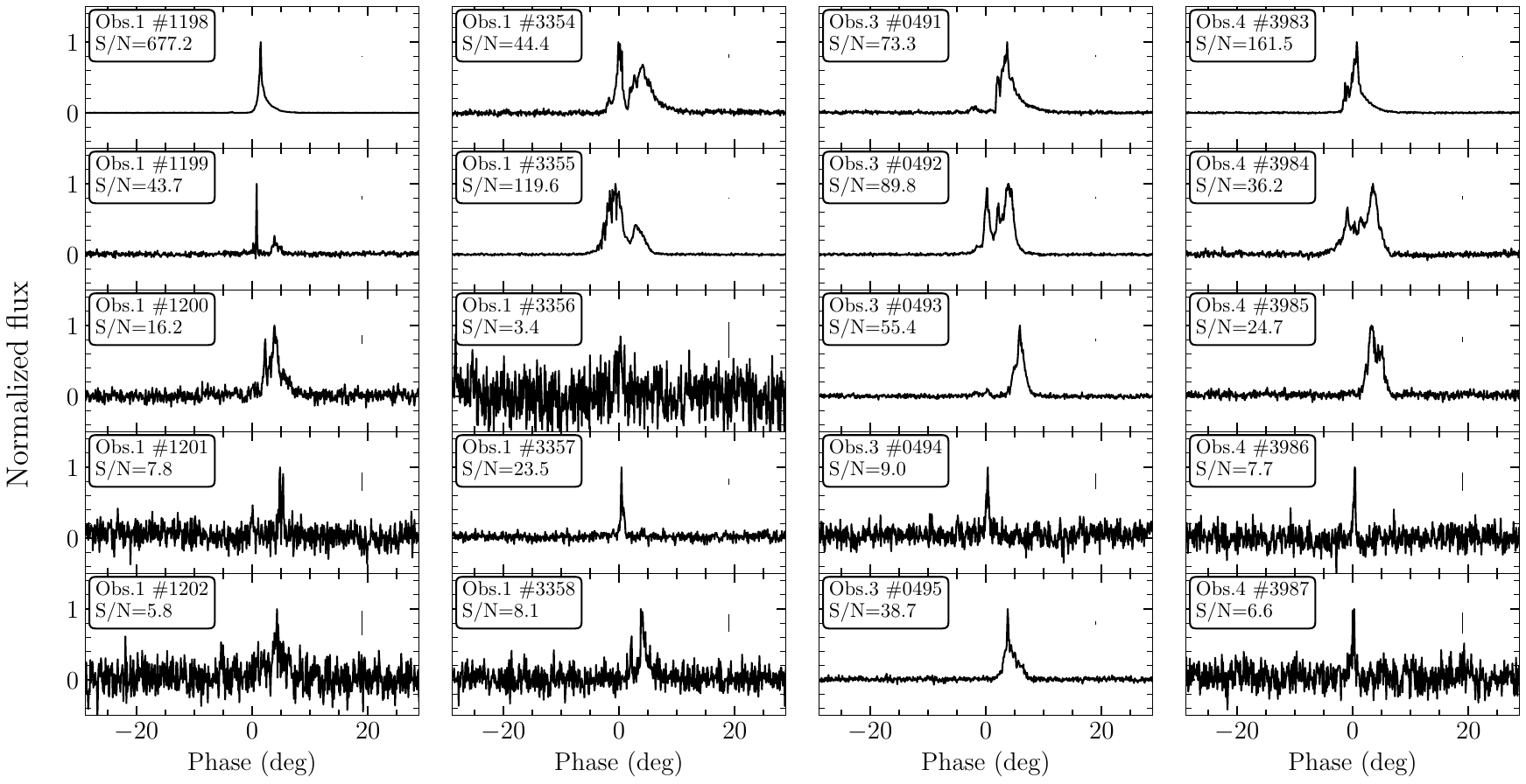}
    \caption{Aligned pulse profiles of four five-pulse groups (rows) from RRAT~J2325--0530. Each panel shows the normalized flux versus pulse phase for individual bursts within a group (columns), labeled by observation and pulse ID. The peak signal-to-noise ratio of each burst is indicated as S/N=$I_{\rm max}/\sigma_{\rm off}$. The bin size and the root-mean-square ($\sigma_{\rm off}$) of the off-pulse region are indicated by the error bar in the upper right region.\label{fig:5}}
\end{figure*}

Four five-pulse groups were identified in our observations. \autoref{fig:5} shows the aligned pulse profiles of these groups, with each row corresponding to a different group. Within each group, individual bursts are displayed in columns, showing normalized flux versus pulse phase. Each panel is labeled by observation ID and pulse number, and the peak signal-to-noise ratio is given as S/N=$I_{\rm max}/\sigma_{\rm off}$ (using 4096 phase bins). As seen in \autoref{fig:5}, S/N generally decreases over time within each burst group, particularly for Obs.4~\#3983--3987 and Obs.1~\#1198--1202. In these burst groups, S/N varies by two orders of magnitude. For example, Obs.1~\#1198 has an S/N$=677.2$, while pulse Obs.1~\#1202 drops as low as $5.8$.

\bibliography{ref.bib}{}
\bibliographystyle{aasjournalv7}

\end{document}